\documentclass[twocolumn]{aastex63}
\usepackage{soul}

\usepackage{color}

\usepackage{hyperref}
\hypersetup{
 colorlinks,
 citecolor=blue,
 linkcolor=red,
 urlcolor=blue}

\received{September 02, 2020}
\revised{November 20, 2020}
\accepted{December 14, 2020}

\submitjournal{ApJL}

\shorttitle{Homologous flaring activity over a sunspot light bridge}
\shortauthors{Louis \& Thalmann}

\begin{document}

\title{Homologous flaring activity over a sunspot light bridge in an emerging active region}

\correspondingauthor{Rohan Eugene Louis}
\email{rlouis@prl.res.in}

\author[0000-0001-5963-8293]{Rohan Eugene Louis}
\affiliation{Udaipur Solar Observatory, Physical Research Laboratory, Dewali, Badi Road, Udaipur -- 313001, India}

\author[0000-0001-8985-2549]{Julia K. Thalmann}
\affiliation{University of Graz, Institute of Physics/IGAM, Universit\"{a}tsplatz 5, A-8010 Graz, Austria}

\begin{abstract}
Sunspot light bridges are known to exhibit a variety of dynamic and persistent 
phenomena such as surges, small-scale jets etc. in the chromosphere 
and transition region. While it has generally been proposed that magnetic reconnection
is responsible for this small-scale dynamism, persistent flaring activity lasting several hours from 
the same spatial location on a sunspot light bridge, has rarely been reported. We combine 
observations from the Atmospheric Imaging Assembly and the Helioseismic Magnetic Imager 
on board the Solar Dynamics Observatory to investigate homologous flaring activity over 
a small sunspot light bridge in an emerging flux region. The homologous flares all produced
broad, collimated jets, including a B6.4 class flare. The jets rise at a speed of about 
200~km~s$^{-1}$, reach projected heights of about 98~Mm, and emerge from the same spatial 
location for nearly 14~hrs, after which they cease completely. A non-linear force free 
extrapolation of the photospheric magnetic field shows a low-lying flux rope 
connecting the light bridge to a remote opposite-polarity network. 
The persistent flares occur as a result of the rapid horizontal motion 
of the leading sunspot that causes the relatively vertical magnetic fields in the adjacent umbra 
to reconnect with the low-lying flux rope in the light bridge. Our results indicate that the 
flaring ceases once the flux rope has lost sufficient twist through repeated reconnections. 
\end{abstract}


\section{Introduction} 
\label{intro}
Sunspots and pores often comprise bright structures called light bridges (LBs) that 
divide the umbra into two or several, smaller umbral cores. 
LBs are either manifestations of large-scale magneto-convection 
\citep{2008ApJ...672..684R,2010ApJ...718L..78R,2015ApJ...811..137T} or ``field-free'' regions 
of hot plasma in the ``gappy'' umbral magnetic field 
\citep{1979ApJ...234..333P,1986ApJ...302..809C,
2006A&A...447..343S}. As a consequence, LBs are often seen along fractures where  
sunspots coalesce or disintegrate  
\citep{1987SoPh..112...49G,2010A&A...512L...1S,2012ApJ...755...16L}.

The disruption of the umbral magnetic field by the LB forces the adjacent umbral magnetic field 
to form a canopy \citep{2006A&A...453.1079J}, which has been suggested to facilitate magnetic
reconnection in the form of several transient phenomena in the chromosphere and transition region. 
These include surges 
\citep{1973SoPh...28...95R,2001ApJ...555L..65A,2016A&A...590A..57R,2015ApJ...811..137T}, 
strong arc-like or extended brightenings \citep{2008SoPh..252...43L,
2009ApJ...696L..66S,2009ApJ...704L..29L}, reconnection jets 
\citep{2014A&A...567A..96L,
2016AN....337.1033L,2017A&A...597A.127B,2018ApJ...854...92T,2019ApJ...870...90B},  
small-scale transient brightenings which excite chromospheric oscillations in the adjacent 
umbra \citep{2017ApJ...850L..33S}, chromospheric shocks
\citep{2017ApJ...835..240S,2019ApJ...886...64Y}, and oscillatory surge-like 
phenomena \citep{2017ApJ...838....2Z,2017ApJ...848L...9H}.

The small-scale jets originating from LBs reported so far 
exhibited characteristic plasma temperatures below 0.1~MK 
\citep{2016A&A...590A..57R,2018ApJ...854...92T} and the general notion 
is that the jets occur as a result of magnetic reconnection between the overlying umbral 
field with either emerging bipoles in the LB or with the twisted flux tube comprising the 
LB \citep{2009ApJ...696L..66S}. However this is at odds with most spectropolarimetric observations
with exceptions as seen in \cite{2015AdSpR..56.2305L}.  
\cite{2015A&A...584A...1L} reported the emergence of a small-scale, flat $\Omega$-loop in a LB.
However, this loop only produced a temperature excess of about 700~K in the chromosphere 
with close to non-zero velocities, when it encountered the overlying umbral field, thus ruling
out magnetic reconnection arising from the emergence. The information provided by the photospheric
magnetic field does not unambiguously explain how, and if, magnetic reconnection in the LB can
produce the recurrent, small-scale phenomena seen in the chromosphere and transition region. 

In this article we study the homologous flaring activity in a sunspot LB 
during the early emerging phase of active region (AR) NOAA 11515, which involved the splitting of 
the AR's main sunspot. These strong, large-scale, homologous flares in a LB, 
which included a B6.4 flare (SOL2012-07-01T01:27), are observed rarely and we provide 
explanations for this particular activity pattern, based on coronal 
field modeling, for the first time here.

\begin{figure}[!ht]
\centerline{
\includegraphics[angle=90,width = 1.02\columnwidth]{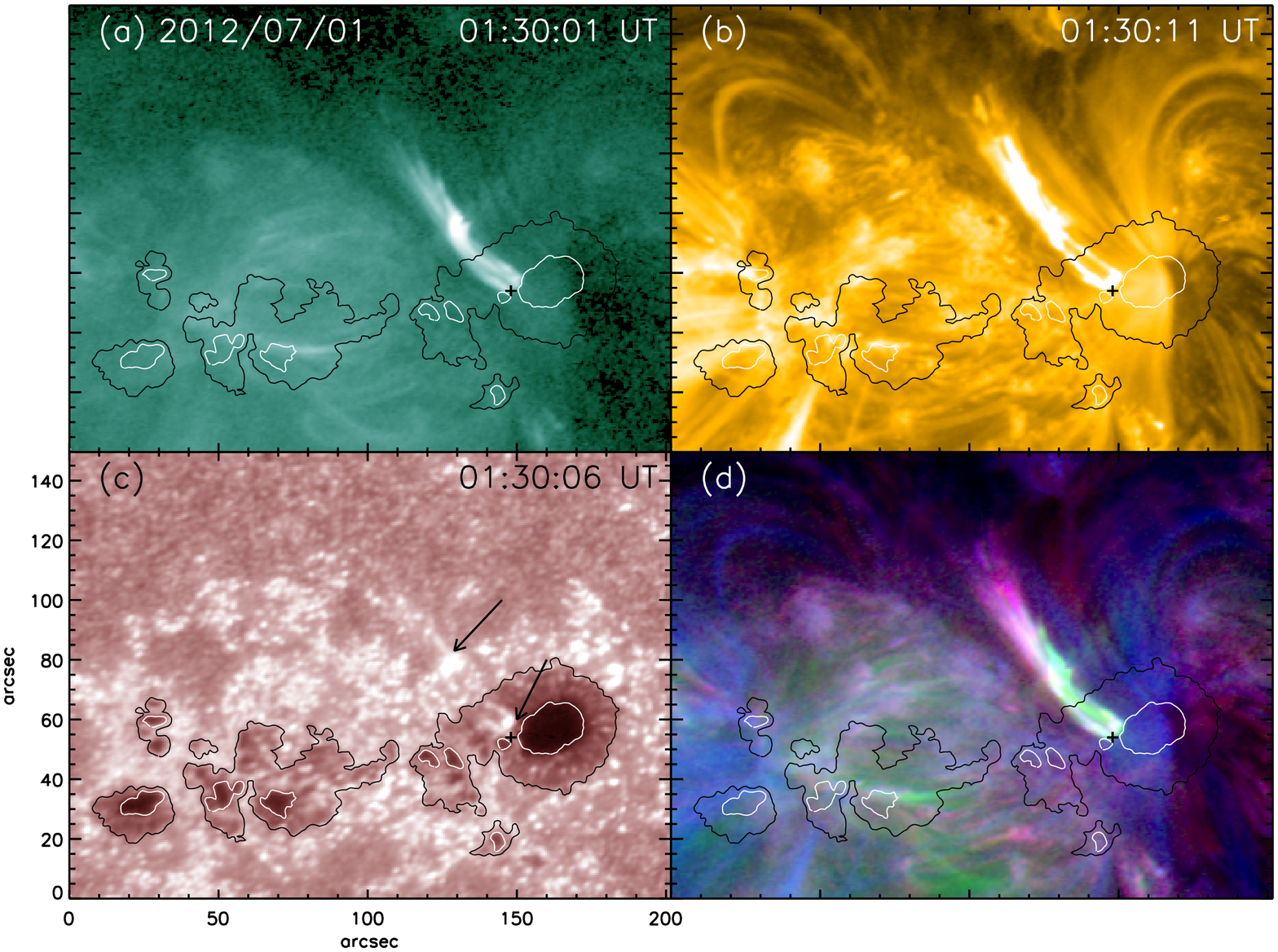}
}
\vspace{-5pt}
\centerline{
\hspace{-15pt}
\includegraphics[angle=90,width = 1.05\columnwidth]{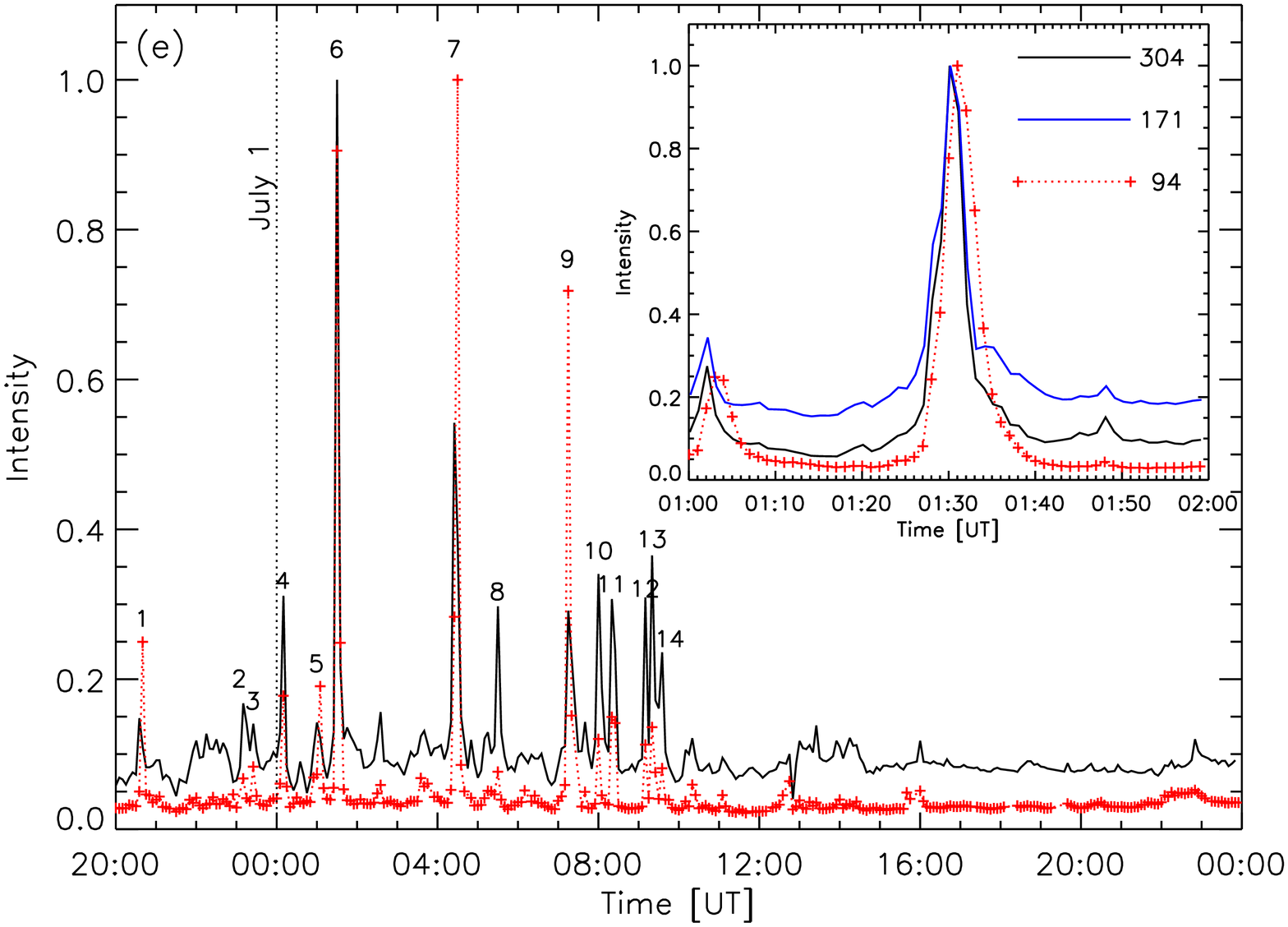}
}
\vspace{-5pt}
\caption{B6.4 flare in the leading sunspot of NOAA AR 11515. 
Panels a, b, and c correspond to AIA 94~\AA, 171~\AA~and 1700~\AA~channels, respectively. 
Panel d is a color composite image made from the AIA 304~\AA~(red), 94~\AA~(green), and 171~\AA~(blue) 
channels.  The black {\textit plus} symbol indicates the base of the jet on the LB. 
Panel e shows the light curve extracted from the mean intensity within the contour outlining 
the jet. Black and red curves correspond to AIA 304~\AA~and 94~\AA~channels, respectively. The 
inset shows the AIA light curve around the time of the B6.4 flare.}
\label{fig01}
\end{figure}

\begin{figure*}[!ht]
\centerline{
\includegraphics[angle=90,width = 1.02\textwidth]{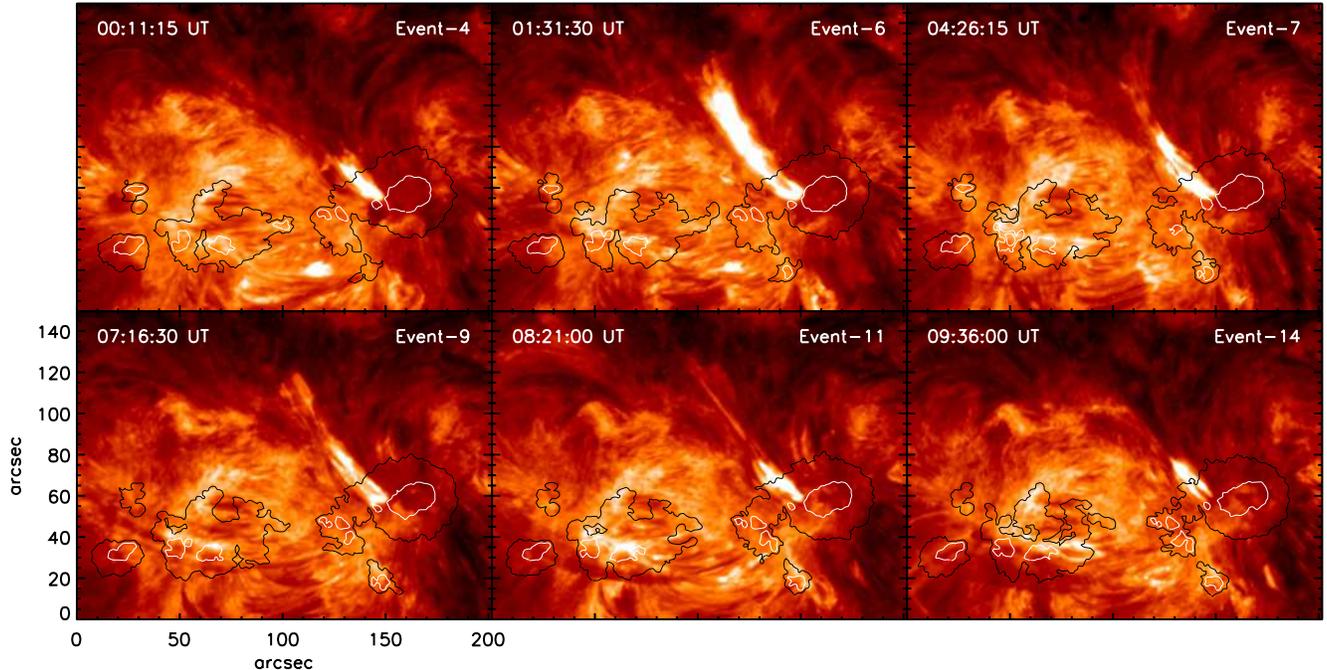}
}
\vspace{-20pt}
\caption{Examples of flaring activity on the LB in the leading sunspot on 
2012 July 01 seen in the AIA 304~\AA~images. See accompanying animation.} 
\label{fig02}
\end{figure*}

\section{Observations and Analysis}
\label{obs}
\subsection{Data}
\label{data}
We use data products from the Helioseismic and Magnetic Imager 
\citep[HMI;][]{2012SoPh..275..229S} and the Atmospheric Imaging Assembly \citep[AIA;][]{2012SoPh..275...17L} 
on board the Solar Dynamics Observatory \citep[{\it{SDO}};][]{2012SoPh..275....3P}, 
to study the time evolution of NOAA AR~11515 from 2012 June~30, 20:00~UT to 2012 July~2, 00:00~UT.
The HMI continuum images and Dopplergrams  have a cadence of 45~seconds.  
The AIA data comprise images in the 1700\,\AA, 304\,\AA, 171\,\AA, and 94\,\AA~ channels with a 5\,min cadence.
Projection effects are assumed to be negligible as the heliocentric angle of the AR was about 33$^{\circ}$.

In order to study the three-dimensional (3D) coronal magnetic field in and around AR~11515, we use 
{\sc hmi.sharp\_cea\_720s} data which provides the Cylindrical Equal Area (CEA) projected photospheric 
magnetic field vector within automatically-identified AR 
patches \citep[][]{2014SoPh..289.3549B} with the azimuthal component of the vector magnetic field being
disambiguated \citep[][]{1994SoPh..155..235M,2009SoPh..260...83L}. We use the full-resolution CEA data 
(pixel scale $\approx$~360~km at disk center) at 2012 July 1, 01:36~UT and 17:00~UT as input to a 
nonlinear force-free (NLFF) method (see Sect~\ref{model}).

\subsection{Modeling}
\label{model}
In order to perform the NLFF modeling, we apply the method of \cite{2012SoPh..281...37W}, i.e., 
we combine the improved optimization scheme of \cite{2010A&A...516A.107W} and a multi-scale approach 
\citep[][]{2008JGRA..113.3S02W} to the 
preprocessed vector magnetic field data. The latter is achieved by applying the preprocessing method 
of \cite{2006SoPh..233..215W} to the original vector magnetic field data. For both, preprocessing and 
optimization, standard model parameter choices, as suggested in \cite{2012SoPh..281...37W}, have 
been used. We adopted a computational domain of $544\times336\times432~\mathrm{pixel}^3$, with the 
photospheric magnetic flux on the lower boundary (at $z=0$) being balanced to within $\sim10\%$.

Successful NLFF modeling is expected to deliver a 3D corona-like model magnetic field with a 
vanishing Lorentz force and divergence. In order to quantify the force-free consistency we use 
the current-weighted angle between the modeled magnetic field and electric current density 
\citep[][]{2006SoPh..235..161S} and find characteristic values of $\simeq10^\circ$. In order to 
quantify the divergence-free consistency of our NLFF solutions we compute the fractional flux 
metric as introduced by \cite{2000ApJ...540.1150W} and find values of $\simeq1.5\times10^{-4}$.

\begin{figure}[!h]
\centerline{
\hspace{20pt}
\includegraphics[angle=90,width = \columnwidth]{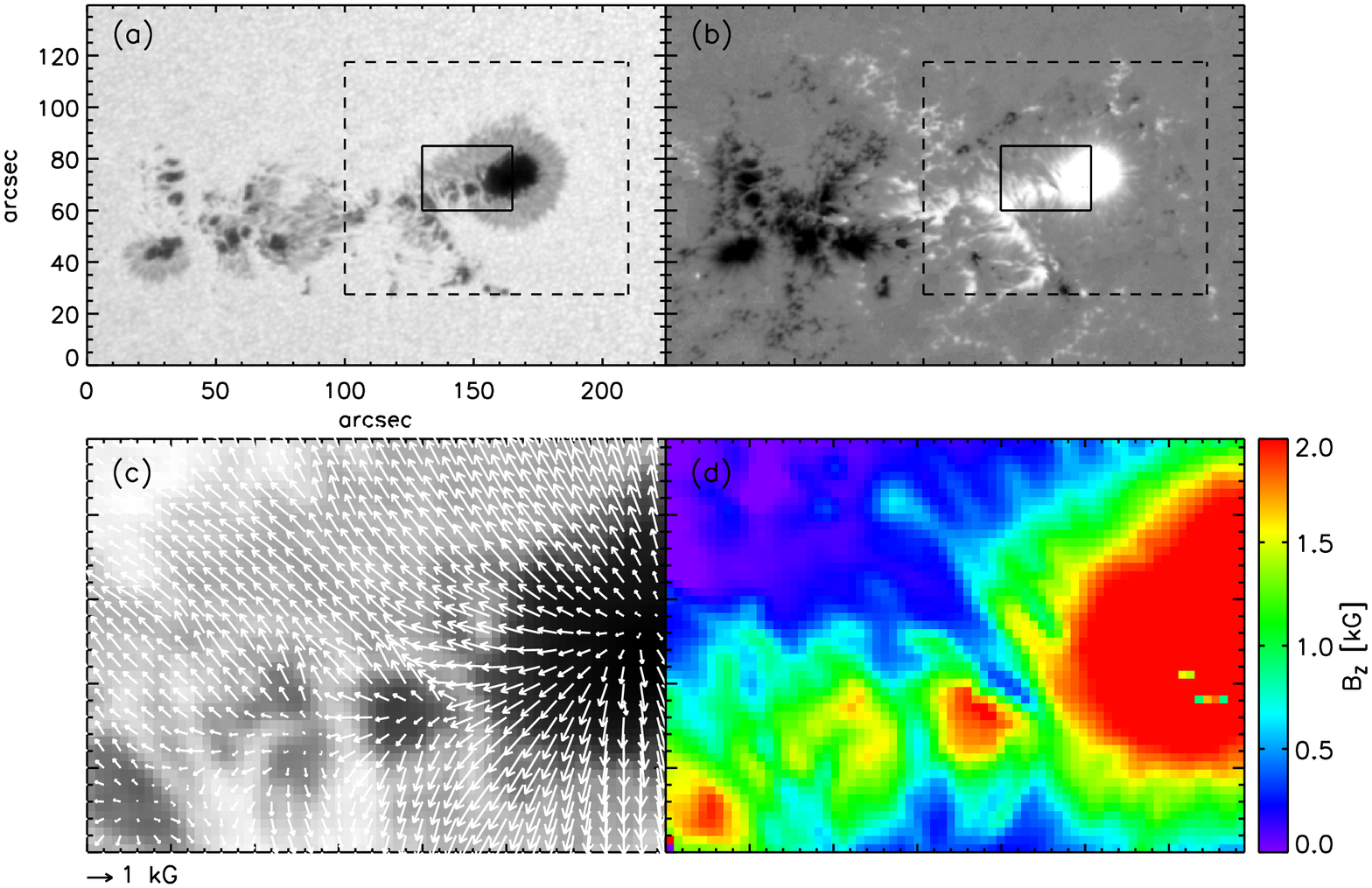}
}
\vspace{-30pt}
\centerline{
\includegraphics[angle=0,width = \columnwidth]{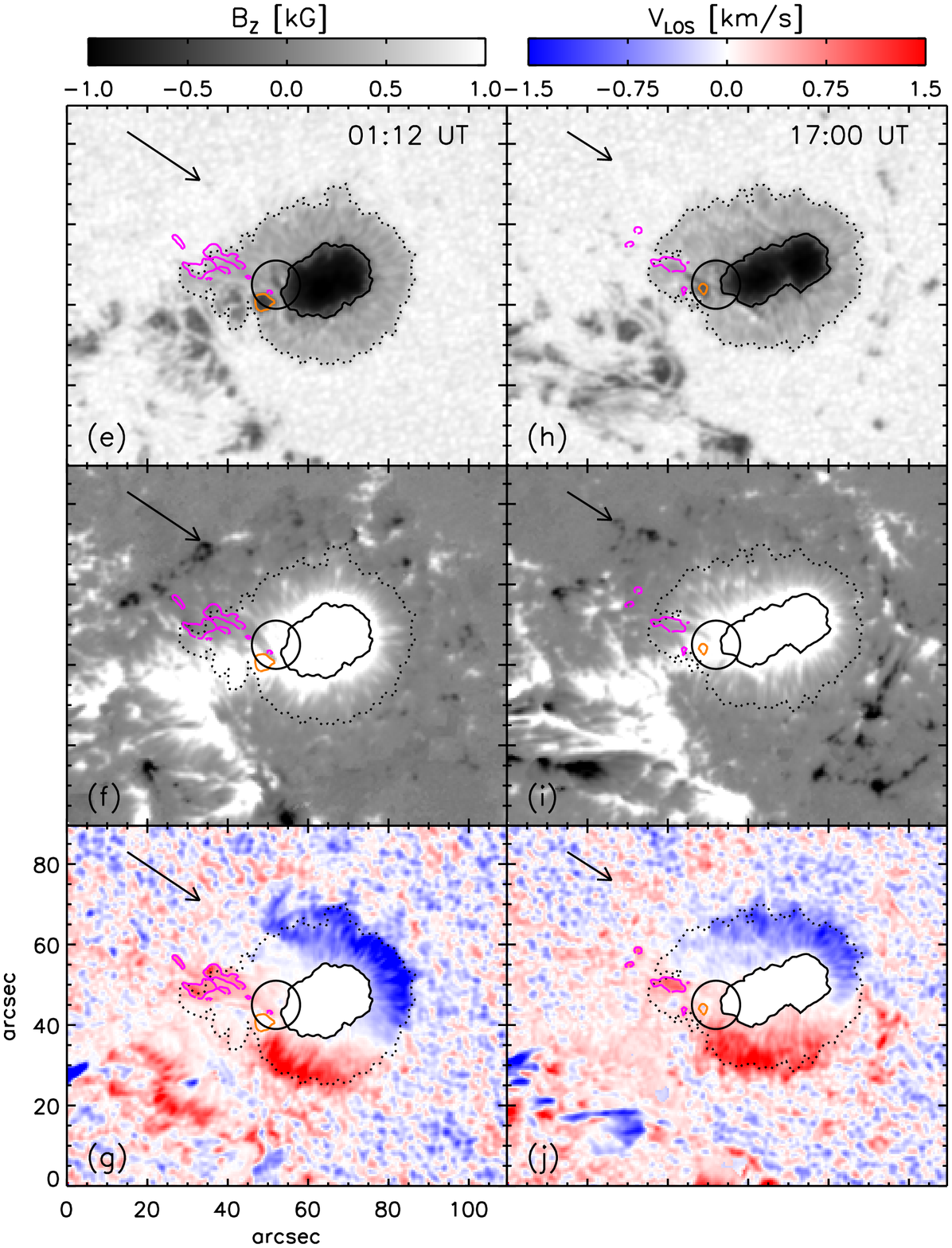}
}
\vspace{-30pt}
\caption{Top: Continuum image (left) and vertical magnetic field (right) 
on 2012 July~01 at 01:12~UT. Middle: Enlarged FOV (indicated by the black box in panel a) of 
the LB in the eastern part of the leading sunspot. Panel c corresponds to
the horizontal magnetic field, shown with white arrows for every second pixel. Panels e--g correspond to the 
continuum intensity, vertical component of the magnetic field, and line-of-sight (LOS) velocity at 01:12~UT. 
Panels h--j are the same as those on the left but at 17:00~UT. The magenta contours 
indicate red-shifts of 550~m~s$^{-1}$ and the black arrows point to a network flux patch. 
The thick black circle encloses the LB. The dotted black contours outline the 
sunspot boundary, while the orange and black solid contours denote the smaller and larger 
umbral cores, respectively.}
\label{fig03}
\end{figure}

\section{Results}
\label{result}
\subsection{Homologous Flaring Activity}
\label{flare}
During its early phase of emergence and 
evolution, between July 01 and July 02, the leading sunspot of AR 11515 rapidly 
split into two, nearly equal, halves with the front half separating from its rear 
twin. The M5.6 (SOL2012-07-02T10:43) and C8.3 (SOl2012-07-01T15:41) flare/CME events 
associated with the splitting of the sunspot have been studied in detail by 
\cite{2014A&A...562A.110L} and \cite{2015SoPh..290.3641L}, respectively. 
In our work, we present a detailed analysis of
the homologous flaring activity on 2012 July~01 that originated from a small 
LB located in the leading positive-polarity sunspot prior to the splitting.

Figure~\ref{fig01}a shows the coronal emission in the AR on 2012 July~01 at 01:30~UT, 
when a B6.4 flare originated from a thin LB (location marked by a black cross) that 
divided the main sunspot into two umbral cores. The related emission is observed in 
the form of a rather broad collimated jet, in the absence
of classical flare ribbon emission
\cite[for a corresponding observation see, e.g.,][]{2003ApJ...589L.117B}.
The base of the jet coincides with the location of the LB. 
Discrete blobs are seen in the 1700~\AA~image, at the base and mid-section of the jet (black arrows).
A color composite image (panel d) shows that the central section in the lower half of the 
jet comprises plasma at coronal temperatures while, the upper half exhibits 
chromospheric/transition region temperatures.  

Fig.~\ref{fig01}e shows the normalized mean intensity at characteristic 
chromospheric and transition region (AIA~304~\AA; $\sim$0.05~MK)
as well as coronal (AIA~94~\AA; $\sim$6~MK) temperatures, computed for all pixels within 
a contour outlining the maximal extent of the B6.4 flare related emission. 
There were as many as 14 flaring events between June 30 20:00~UT and July 01 10:00~UT, that were 
observationally verified, with the B6.4 flare being the strongest event. The flare activity in 
the LB ceased after around 12:00~UT on July 1st.

Figure~\ref{fig02} shows the morphology of selected flares emanating from the LB.  
From the time series of observations, an untwisting motion
is recognized, with the plasma descending along similar trajectories during the individual 
events. The plasma comprising the jet moves outwards with a maximum projected speed 
of about 200~km~s$^{-1}$ with the cool plasma being ejected to a projected height 
of about 98~Mm from the base of the jet. These representative values were estimated 
from the 304~\AA~channel, using a space-time map along a parabolic cut outlining the ejection 
for Event 9 at 07:16\,UT.

The homologous flaring activity described above is seen in all the AIA channels,
including the 94\,\AA~channel. The events are also detected in the GOES X-ray flux curve, at 
least for the B6.3 class flare, while the majority of other events are weaker, B-class flares 
(comparing the relative intensity in the 94\,\AA~channel of the B6.3 flare). In comparison,
the small-scale jets/surges reported previously in LBs, are confined to the 
lower chromosphere to transition region, where their temperatures do not exceed 0.1\,MK, 
and clearly lack the intense emission associated with the flares (Figs.~\ref{fig01} and \ref{fig02}). 
Additionally, the spatial scales of the flares are far larger than those of small-scale 
jets/surges which are typically ejected to 10--15\,Mm. These characteristics make the flares 
in the LB distinct from the recurrent jets and surges reported previously.

\subsection{Photospheric Magnetic Field Configuration}
\label{magnetic}

Figure~\ref{fig03} shows the photospheric structure of AR~11515 at 01:12~UT (top row). 
The field strength in the umbral cores on either side of the LB is about 2~kG while the average and 
minimum field strength in the LB is about 1.2~kG and 1.0~kG, respectively. 
The field inclination is about 60--70$^\circ$ in the LB and about 80$^\circ$ in the penumbra. 
However at HMI's spatial resolution we do not see any indication of the field changing polarity 
either in the LB or in the penumbral filament extending into the LB (panel d).


The bottom panels of Fig.~\ref{fig03} show the leading sunspot at 2 different instances of time 
on 2012 July~01, before the B6.4 flare at 01:12~UT (panels e--g) and after the jet/flaring activity ceased
at 17:00~UT (panels h--j). Evidently, by 17:00~UT, the LB  has become more diffuse 
and the penetration of the highly inclined field, associated with the LB, into the umbra has 
receded significantly. The area and hence the magnetic flux of the smaller umbral core east of 
the LB (orange contour), also reduces by 75\%, from 1.5$\times 10^{20}$~Mx to 3.6$\times 10^{19}$~Mx. 
In comparison, the flux of the larger umbral core only reduces by about 5\%. 

The LB as well as the associated penumbral sector harbor red-shifts
of up to 550~m~s$^{-1}$ at 01:12~UT, which become considerably smaller at 17:00~UT. 
A comparison of Figs.~\ref{fig03}f and \ref{fig03}i show signatures of flux cancellation at 
the edge of the moat, north of the main sunspot possibly due to the relative motion of the 
leading sunspot in a north-westerly direction. 

\begin{figure}[!h]
\centerline{
\includegraphics[angle=0,width = \columnwidth]{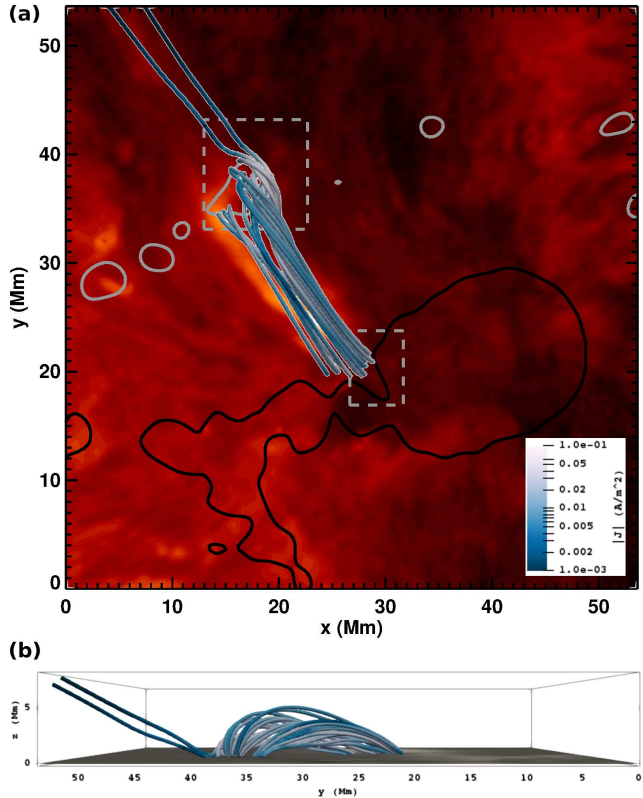}
}
\vspace{-5pt}
\caption{
NLFF modeling of the coronal magnetic field of AR~11515 on 2012 July~1 at $\sim$01:36~UT. Shown is a sub-domain of 
the original NLFF model volume, covering the connectivity of the observed LB located 
within the positive-polarity leading sunspot (see dashed outline centered around $(x,y)=(28,21)$~Mm)
and the moat area to the north-east of the AR (see dashed outline centered around $(x,y)=(15,38)$~Mm).
Field lines rooted with at least one of their footpoints within the LB or moat area and 
with an inherent twist of $\gtrsim -0.35$~turns are shown, 
and color-coded according to the total current along them.
The color-coded background shows the coronal emission in the 304~\AA\ band at 01:35:30~UT. 
Black (gray) contours outline the NLFF lower boundary vertical magnetic field at
1000 (-200)~G, respectively.}
\label{fig04}
\end{figure}

\subsection{Coronal Magnetic Field Configuration}
\label{extrapol}

The NLFF modeling of the 3D coronal magnetic field on July~1 at 01:36~UT is shown in Fig.~\ref{fig04}. 
A system of low-lying twisted magnetic field (likely comprising a magnetic flux rope, and so-called
hereafter) connects the LB (outlined by the dashed rectangle centered around $(x,y)=(28,21)$~Mm) 
to the negative-polarity moat area (see dashed outline centered around 
$(x,y)=(15,38)$~Mm and black arrow in Fig.~\ref{fig03}f). 
The particular field lines shown have been selected based on their inherent twist,
with a twist of more than $0.35$~turns (see Sect.~\ref{twist}), and which are
rooted with one of their footpoints within the LB or moat area. The field lines extending beyond the FOV
suggest the presence of bald batches, i.e., locations where the field overlying a polarity inversion line 
is parallel to the photosphere \citep[see, e.g.,][]{1993A&A...276..564T}. The apexes of the twisted field 
lines reach coronal heights up to $\sim5$~Mm (Fig.~\ref{fig04}b). 

Visual comparison to the AIA~304\AA\ image, taken near-simultaneously (color coded background in Fig.~\ref{fig04}a), 
evidences the realistic approximation of the solar corona by the employed NLFF modeling. We find that the direction 
of the jet feature coincides spatially between the LB and moat areas and the extension of the jet feature 
to the north-east of the AR is also well recovered by the field lines associated with the bald patches.

\subsection{Time Evolution of Magnetic Twist}
\label{twist}

We perform an analysis of the inherent magnetic twist (thus self-helicity) of the system of twisted field lines 
shown in Fig.~\ref{fig04}, at 01:36~UT and 17:00~UT.
We consider field lines that are rooted within the LB or moat area (dashed outlines
in Fig.~\ref{fig04}) with at least one of their footpoints, and which emerge from locations where $|B_z|>100$~G,  
at the NLFF lower boundary. Based on these criteria, we compute 167 (104) field lines at 01:36~UT (17:00~UT).

\begin{figure*}[!ht]
\centerline{
\includegraphics[angle=0,width =\textwidth]{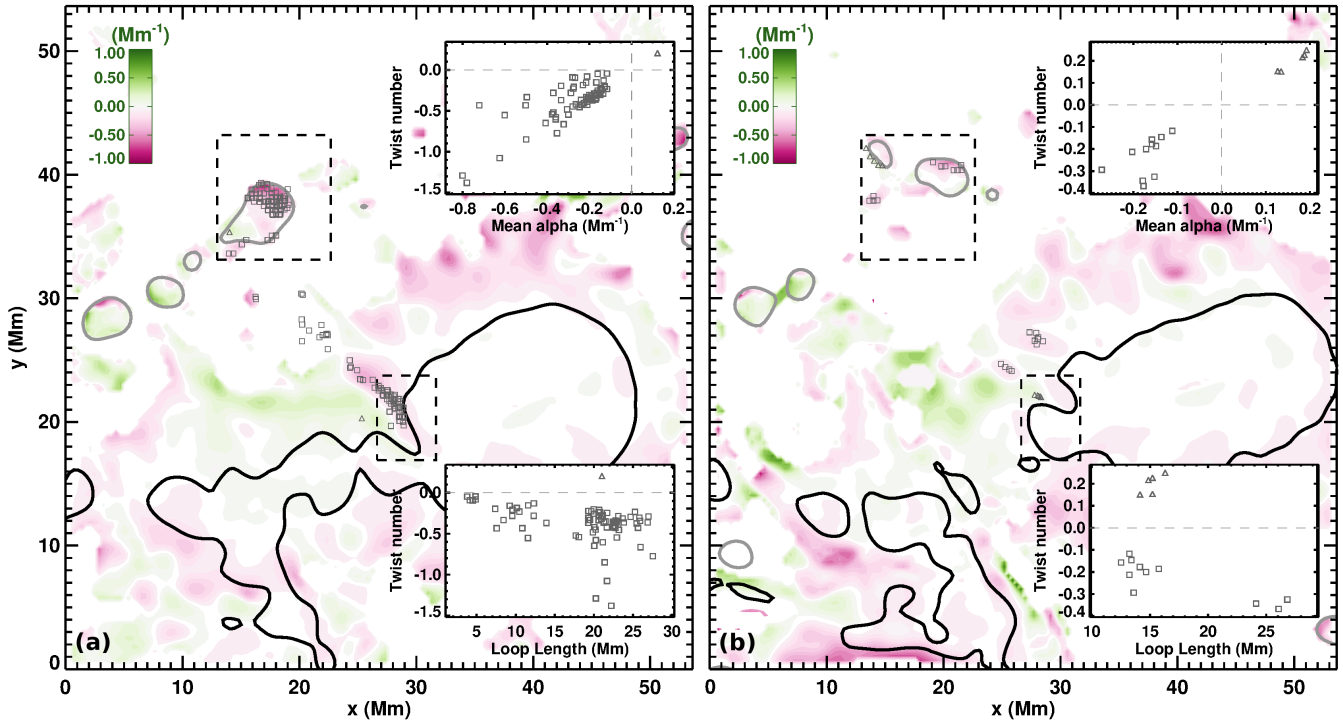}
}
\vspace{0pt}
\caption{
Force-free $\alpha$  and twist parameter $T_{\overline{\alpha}}$ 
within NOAA~ 11515 at (a) 01:36~UT and (b) 17:00~UT 
on 2012 July~01. Dashed outlines to the north-east ($x,y=15,38$~Mm) and south-west ($x,y=28,21$~Mm) outline the moat 
area and LB, respectively from which field lines were computed. Footpoint locations of all considered field lines 
adhering to positive 
(negative) $T_{\overline{\alpha}}$ are marked by gray triangles (squares).  The color-coded background shows the 
force-free parameter $\alpha$, scaled to $\pm1~\rm{Mm}^{-1}$. Black (gray) contours outline $B_z$ of $1000$~$(-200)$~G.
The insets show the field lines' twist as a function of $\alpha$ (top right corners) and 
as a function of the field lines' lengths (bottom right corners).}
\label{fig05}
\end{figure*}

Subsequently, we estimate the twist for all computed field lines as
$T_{\overline{\alpha}}=\overline{\alpha}L/4\pi$, where $\overline\alpha$ is the mean value of $\alpha$ at 
both footpoints and $L$ is the arc-length of each considered field line \citep[e.g.,][]{lea_can_03}. 
We restrict ourselves to estimate $T_{\overline{\alpha}}$ for all field lines that are rooted at locations 
for which $|\overline{\alpha}|\ge0.1$~Mm$^{-1}$. We set the threshold on $\alpha$ in order to avoid that
the twist estimation is not dominated by the many existing small-scale field lines.
This reduces the number of analyzed field lines to 74 (16) at 01:36~UT (17:00~UT). 
The footpoints of these remaining field lines are indicated by gray triangles (positive twist) and squares 
(negative twist) in Fig.~\ref{fig05}. The force-free parameter $\alpha=\mu_0 J_z/B_z$, where $J_z$ is the 
vertical component of the electric current density, is displayed as a color-coded background in Fig.~\ref{fig05}. 

We find that at 01:36~UT the majority of the considered field lines rooted in the LB and moat areas,
are linked to locations of negative $\alpha$, and consequently adhere to a negative twist
(see inset at top right corner of Fig.~\ref{fig05}a).  
The distribution of $T_{\overline{\alpha}}$ as a function of $L$ (bottom right 
inset), shows that the negative twist is characteristic for both, shorter and longer field lines. 
From the considered field lines we find a median value of $\langle T_{\overline{\alpha}}\rangle=-0.3\pm0.1$.

At 17:00~UT the situation is clearly different. The LB appears less pronounced, while the moat region 
moved towards the solar west quite considerably.  
Still, most of the considered field lines are rooted in locations of predominantly
negative $\alpha$, resulting in a median value of $\langle T_{\overline{\alpha}}\rangle=-0.2\pm0.1$. 
Note that the number of considered field lines is greatly reduced, by a factor of nearly 
five, as many more of the field lines rooted in the selected sub-fields correspond to a value of 
$|\overline{\alpha}|<0.1$~Mm$^{-1}$, based on the criterion described above.

These results suggest that the flux rope connecting the LB and moat area lost a considerable amount of
twist, as the number of field lines with a twist $\gtrsim 0.1$ at 01:36 UT, reduced 
significantly at 17:00 UT, likely due to the repeated reconnection events.

\section{Discussion}
\label{discuss}

The homologous flaring activity in the sunspot LB lasts for at least 14~hrs with the strongest 
event being a B6.4 flare (see Fig.~\ref{fig02}). 
In the present case, only the 3D NLFF reconstruction of the coronal magnetic field reveal the 
importance of temporal changes within specific photospheric areas of the AR, such as the moat region 
located to the north-east of the main sunspot (see Fig.~\ref{fig04}). In particular, it is connected by 
a low-lying magnetic flux rope to the LB, spatially associated to the observed flare events.

The chief driver for the recurrent flares originating from the LB is the horizontal motion of the leading 
sunspot in a north-westerly direction. Consequently the more vertically oriented magnetic fields in the 
adjacent sunspot umbra to the east of the LB, may reconnect with the highly inclined fields at the leg of 
the flux rope producing the observed collimated jet emission (top panel of Fig.~\ref{fig04}). During 
the several subsequent, homologous reconnection events the flux rope's twist gets liberated (Fig.~\ref{fig05}), 
resulting in a weakly twisted/sheared field arcade no longer conducive to magnetic reconnection. 
Additionally, flux cancellation at the remote leg of the flux rope in the moat area could facilitate reconnection 
in the rope, and the flaring ceases as the footpoint diffuses. The reduction of photospheric magnetic 
flux at/close to the regions where the eruptions ensue is generally consistent with observations and MHD simulations 
\citep[][and references therein]{2006SSRv..123..251F}.

The LOS velocity in the LB and its surroundings is dominantly red-shifted (Fig.~\ref{fig03}g). 
If the homologous nature of the flares
were the result of small-scale loops emerging in the LB and reconnecting with the overlying umbral field
\citep{2018ApJ...854...92T}, one would expect blue-shifts either in localized patches or along 
the body of the LB \citep{Louisetal2020}. With HMI's native spatial resolution of $\sim$320~km, 
we do not find evidence for opposite-polarity fields above uncertainties within the LB, though we 
do not exclude the possibility of small-scale velocity and magnetic inhomogeneities having been present
\citep{2009ApJ...704L..29L,2015A&A...584A...1L,2015AdSpR..56.2305L}.

It has been known that LBs can be associated with the large-scale magnetic configuration of an AR. Observations 
by \cite{2003ApJ...589L.117B} describe a set of flare ribbons extending along the LB as a result of magnetic 
reconnection in the corona. However, the C2.0 flare was the only event that occurred in the AR on that day. 
\cite{2010ApJ...711.1057G} reported recurrent surges in H$\alpha$ from a LB which triggered a filament eruption 
and an associated M2.5 flare. However, the lack of vector magnetic field data does not account 
for the persistent nature of the surges or the topological association of the filament to the LB. 
The homologous flaring activity in a sunspot LB, described in this article, is observed 
unambiguously in all AIA channels and is strongly supported by our models of the coronal magnetic field, 
which, to the best of our knowledge, has not been reported previously.

\section{Conclusions}
\label{conclude}
The presence of small-scale chromospheric and coronal transients such as surges, jets, etc. in sunspot LBs has  
generally been attributed to magnetic reconnection but the processes, rendering these transients homologous, are 
poorly understood. Homologous flares in LBs can be considered extremely unique and rare, as they are an irrefutable
evidence of magnetic reconnection occurring within the strong field domain of sunspots.
In this article, we describe a series of intermittent flares in a sunspot LB. The leading sunspot exhibits 
a rapid horizontal motion consequently splitting into 
two, nearly equal halves. Our coronal magnetic field models show a low-lying twisted flux rope connecting 
the LB to the moat area of the sunspot. The proper motion of the sunspot causes the umbral magnetic fields to reconnect 
with the low-lying flux rope, producing the flares, with the collimated jets perfectly aligned along the flux rope's axis. 
The repeated reconnections render a considerable loss of twist in the flux rope 
which is substantiated by our extrapolations.  The combined effect of the above renders the flux 
rope to return to a more relaxed state whereby the flaring activity ceases. The uniqueness of the observations and the 
coronal magnetic field models provide new insight into the recurrent nature of flares in a LB.

\acknowledgements
SDO data are courtesy of the NASA/ SDO AIA and HMI science teams. They are provided by the Joint Science 
Operations Center – Science Data Processing at Stanford University. J. K. T. was supported by the Austrian 
Science Fund (FWF): P31413-N27. We thank the referee for providing useful comments.


\end{document}